\documentclass{aa}  

\usepackage{graphicx}
\usepackage{txfonts}
\usepackage{color}
\usepackage{url}
\usepackage{hyperref}

\usepackage{upgreek}
\usepackage{graphicx}	
\usepackage{amsmath}	
\usepackage{amssymb}	
\usepackage{xcolor}
\usepackage{comment}

\usepackage[normalem]{ulem}
\usepackage{xcolor}
\definecolor{new_color}{rgb}{1.0, 0.43, 0.3}

\newcommand{\uu}[1]{_{_{\rm #1}}}
\newcommand\redsout{\bgroup\markoverwith{\textcolor{red}{\rule[0.5ex]{2pt}{0.4pt}}}\ULon}
\begin{document}

   \title{Constraints on the accretion properties of quasi-periodic erupters from GRMHD simulations}

   \titlerunning{Accretion properties of QPEs}


   \author{A. Chashkina
          \inst{1,2}\fnmsep\thanks{chashkina.anna@gmail.com},
          O. Bromberg\inst{2}, A. Levinson\inst{2}
          \and
          E. Nakar\inst{2}}

   \institute{Max Planck Institute for Astronomy, Königstuhl 17, 69117 Heidelberg, Germany \\
   \and
   The Raymond and Beverly Sackler School of Physics and Astronomy, Tel Aviv University, Tel Aviv 69978, Israel
}


 
  \abstract
   {Some apparently quiescent supermassive black holes (BHs) at centers of galaxies show quasi-periodic eruptions (QPEs) in the X-ray band, the nature of which is still unknown. A possible origin for the eruptions is an accretion disk, however the properties of such disks are restricted by the timescales of reccurance and durations of the flares. }
   {In this work we test the possibility that the known QPEs can be explained by  accretion from a compact accretion disk with an outer radius $r_{\rm out}\sim 10-40 r_{\rm g}$, focusing on a particular object GSN 069.}
   { We run several 3D GRMHD simulations with the {\tt HARMPI} code of thin and thick disks and study how the initial disk parameters such as thickness, magnetic field configuration, magnetization and Kerr parameter affect the observational properties of QPEs.}
   {We show that accretion onto a slowly rotating BH through a small, thick accretion disk with an initially low plasma $\beta$ can explain the observed flare duration, the time between outbursts and the lack of evidence for a variable jet emission. In order to form such a disk the accreting matter should have a low net angular momentum. A potential source 
   for such low angular momentum matter with a quasi periodic feeding mechanism might be a tight binary of wind launching stars. }
   {}




   \maketitle
%

\section{Introduction}
\label{sec:introduction}

Quasi-periodic eruptions (QPE) were detected for the first time in the source GSN 069 in 2019 \citep{Miniutti2019}. During the  active periods the observed X-ray luminosity increases by up to 2 orders of magnitude above its quiescent level, reaching a peak luminosity $L_{x}=10^{42}-10^{43}$~erg~s$^{-1}$. The source has also been detected at $6$~GHz by the VLA and has a persistent radio luminosity $L_{\rm radio}\sim 1.9\cdot 10^{36}\,{\rm erg\, s^{-1}}$, and a corresponding jet power $P_j \approx 10^{41}\,{\rm erg\, s^{-1}}$  \citep{Miniutti2019}. Later four other sources were found -- RX J1301.9+2747, eRO-QPE1, eRO-QPE2, and XMMSL1 J024916.6$-$041244 \citep{Giustini2020, Arcodia2021, Chakraborty2021}. All of them have similar variability patterns: short X-ray outbursts (0.4 - 8 hours) separated by long quiescent states (2-19 hours). The durations of the outbursts, their amplitudes, and the recurrence times are not constant for a given source, but change from burst to burst in such a way that the strongest QPE are followed by longer recurrence times \citep{Miniutti2019}. The spectra in both active and quiescent states are soft and well fitted by a black-body with a temperature of $kT\sim 30-50$~eV in the quiescent and $kT\sim 100-300$~eV during the active states, although quiescent emission has not been detected from all sources. Assuming that the quiescent black body emission comes from the disk inner regions, one may estimate the black hole masses in the observed QPEs to be in the range  $M_{\rm BH}=10^5-{\rm few}\times 10^6M_{\odot}$. All sources are associated with galactic nuclei, but their masses are smaller than typical BH masses in AGNs.  Another feature distinguishing these sources from classical AGNs is the lack of broad optical or UV emission lines, although in GSN 069 there are narrow line regions that might indicate recent activity.  According to \citet{Miniutti2019}, the lack of variability in 
the radio source detected in GSN 069 implies that the production mechanism of the associated jet is likely unrelated to the QPE process. As we argue below, this may not necessarily be the case. The observed parameters of each of the five sources are given in Table~\ref{tab:sources}.  

QPEs are widely discussed in the community \citep[e.g.][]{Pan2022, Quataert2023, Sukova2021, Metzger2022, Ingram2021, Wang2022}. A major challenge for any model involving disk accretion is the viscous time scale, which is determined by the disk size and can easily exceed the observed outbursts duration. One possible solution is that the outbursts originate from the innermost parts of the disk, and are thus disconnected from the overall viscous time. 
For example, \citet{Pan2022} assumed that the QPEs are produced by radiation- or magnetic-pressure-driven instabilities at the inner part of the accretion disk (see also \citet{Sniegowska2023}). 
\citet{Quataert2023} explained the QPEs with circularization shocks {taking place during unstable mass transfer} events onto the BH from a star undergoing a Roche lobe overflow (RLOF). 

 We will discuss in more details two models that are particularly relevant to our work.  In the first, \citet{Metzger2022} considered two stars orbiting a black hole on coplanar orbits, where at least the outer star is undergoing a Roche lobe overflow (RLOF). When the inner star passes between the outer star and the black hole, the combined gravitational pull reduces the Roche lobe radius of the outer star, causing a short-term increase in the mass transfer rate from the outer star. As a result the accretion disk is loaded with fresh stellar material. 
 In this model the disk accretion is the source of the QPE, while the recurrence time is determined by the time interval between flybys of the two stars. In such a system the accretion disk formed during the flybys could be much more compact due to the sideways gravitational pull of the second star, thus resolving the long viscous time problem. 

In the second work \citet{Krolik2022} connected the origin of the QPEs with the formation of a self-interacting shock in the accretion stream from a star orbiting the black hole. In this model it was assumed that a significant amount of the angular momentum of the  matter is carried away by magnetic stresses, so the shocked matter falls down onto the BH within the free-fall time. Existence of magnetic stresses suggests a prompt formation of an accretion disk, in which the stresses are responsible for the effective viscosity and angular momentum redistribution. From simulations of disk accretion, we know that magnetic pressure is limited by approximate equipartition, and the magnetization of the accretion flow itself is never very high. For the \citet{Krolik2022} model to be successful, it is necessary for the accretion disk to accrete most of its mass during the recurrence time, implying again a small disk size. 

 Both of these works explain some observational features of the QPEs such as the burst/recurrence times and the burst amplitudes, however an important question remains: what happens to the matter during its accretion onto the BH? To account for the flare energy, an excess mass of $\Delta M \sim 10^{-7} M_\odot$ must be transferred from the star to the disk in each cycle, corresponding to a mean mass transfer rate of $\Delta M/ T_{\rm rec} \sim 10^{-4} (\Delta M/10^{-7} M_\odot) (T_{\rm rec}/10 \, {\rm hrs})^{-1}\, M_\odot/{\rm yr}$, where $T_{\rm rec}$ is the flare recurrence time. If the accretion time is longer than the flare recurrence time, the mass of the disk will continuously increase until a quasi steady-state is established, in which, on average, the rate at which mass is transferred to the disk is balanced by the rate of accretion onto the black hole. This would lead to a quiescent bolometric luminosity of
 $L_b \approx 10^{42}$ erg s$^{-1}$, which may be consistent with the observed luminosity during minimum states in GSN 069 and RXJ 1301.9+2747, assuming a correction factor of a few between the UV and soft X-ray luminosities, but not with the quiescent luminosity of other QPEs.  Regardless of these observational constraints,
 in both models, the authors argued that it is necessary for the BH to accrete all of the material within the timescale of the QPE cycle, or even the timescale of the flare itself.
\citet{Metzger2022} made 1D semi-analytical calculations and showed that a significant part of the disk material will accrete onto the BH on timescales less than a day (see Fig. 3 of their paper). 
\citet{Krolik2022} did not attempt to model the infalling material beyond the point of self-intersection. A more self-consistent answer to this question can be found using numerical 3D simulations of the accretion flow. 

\begin{table*}
    \centering
    \begin{tabular}{c|c|c|c|c|c|c}
        Source Name & $\tau_{\rm QPE}$, h & $T_{\rm rec}$, h & $M_{\rm BH}/M_{\odot}$ & $\dot M/\dot M_{\rm Edd}$  & $L^{0.5-2 \rm keV}_{\rm max}$, erg s$^{-1}$ & $L^{0.5-2 \rm keV}_{\rm min}$, erg s$^{-1}$  \\
        \hline
        GSN 069$^{[\rm a]}$  & $\sim 1$ & $\sim 8.4-9.2$ & $0.3-4\cdot 10^6$  & $\sim 0.5$   & $5\cdot 10^{42}$ & $<10^{41}$ \\
        \hline
        RX J1301.9+2747$^{[\rm b]}$ & $\sim 0.5$ & $\sim 3.7 - 5.5$ & $0.8-2.8\cdot 10^6$  & $0.14$   & $1.2-1.6\cdot 10^{42}$  & $\geq 5\cdot 10^{40}$ \\
        \hline
        XMMSL1 J024916.6-041244$^{[\rm c]}$ & $\sim 0.35$   & $\sim 2.5$ & $0.85-5\cdot 10^5$   & $0.13$   & $3.4\cdot 10^{41}$ & $1.6\cdot 10^{41}$ \\
        \hline
        eRO-QPE1$^{[\rm d]}$ & $\sim 7.6$   &  $\sim 18.5$  &   &    & $9.4\cdot 10^{42}$   & $<2\cdot 10^{41}$ \\
        \hline
        eRO-QPE2$^{[\rm d]}$ & $\sim 0.4-1$  & $\sim 2.7$  &   &    & $10^{42}$  & $<4\cdot 10^{40}$ \\
        
         \hline
          
    \end{tabular}
    \caption{  The main observational parameters of QPE systems: the name of the source, the duration of outbursts in hours, the recurrence time in hours, mass accretion rate during the outbursts in Eddington units, black hole mass estimated from black body radiation in quicent state ,  maximal and minimal luminosities in the range $0.5-2$~keV. The data is taking from the papers: [a] -- \citet{Miniutti2019}, [b] -- \citet{Giustini2020}, [c] -- \citet{Chakraborty2021} , [d] -- \citet{Arcodia2021}.}
    
    \label{tab:sources}
\end{table*}

In this work we investigate whether the recurrence times and the QPE duration can be explained by a disk accretion onto a BH without invoking additional physical entities. 
This general result has particular applications to the works of \citet{Metzger2022} and \citet{Krolik2022}. In a system with an orbital period $P_{\rm orb}\sim T_{\rm rec}\sim 10$~hours, the accretion disk size would typically be of the order of a hundred gravitational radii $r_g$.  Information about any changes at the outer boundary of the disk will be transferred into the inner parts on viscous times associated with the outer disk radius, $r_{\rm out}$, which is too large to account for the QPE. An obvious way to account for this inconsistency is to assume that the disk is initially much more compact. In this paper we consider accretion from smaller disks with outer radii $r_{\rm out}\sim 10-40r_g$ and thicknesses $h/r\sim 0.1-0.5$.   The viscous times at the outer boundaries of these disks are $(10-50)/\alpha_{0.1}$~hours, where $\alpha_{0.1}=\alpha/0.1$ is the viscosity parameter,   comparable with the recurrence time. These disk sizes are consistent with the observed luminosities and black body temperatures, both in the quiescent and in the active states, meaning that the emitting region should be very compact, comparable to the innermost stable circular orbit size.
In addition to the fact that such disks have much shorter viscous time scales, the small sizes can also explain the lack of broad optical and UV emission lines that are produced in the outer parts of huge AGNs disks. We do not study how such small disks can be formed, but discuss a few possibilities.  

\begin{table*}
    \centering
    \begin{tabular}{c|c|c|c|c|c|c|c|c|c}
        Model& $N_{\rm loops}$ & $r_{\rm in}/r_{\rm g}$ & $r_{\rm max}/r_{\rm g}$& $a$ & $\beta_{\rm min}$ & $\Psi_0$  & $H/R$ &  $M_{10}/M_0$  \\
        \hline
        {\tt \href{https://www.youtube.com/watch?v=ihz74wVz1Ew}{N1}} & $1$ & $4$ & $7$ & $0.9$ &  $100$ &  $0.01$  & $0.13$  &  $0.84$\\
         \hline
              {\tt \href{https://youtu.be/JbfZfpy069M}{N2}}& $1$ & $8$ & $10$ & $0.9$ &  $100$ & $0.03$ & $0.13$ &  $0.93$\\
         \hline
          {\tt \href{https://youtu.be/uXf-CR7HsHI}{K1}} & $1$ & $8$ & $14$ & $0.9$ &  $1000$ & $0.18$  & $0.46$ &  $0.89$ \\
         \hline
          {\tt \href{https://youtu.be/28BXgrisLYM}{K2}} & $6$ & $8$ & $14$ & $0.9$ &  $100$ & $0.23$ & $0.46$ &  $0.97$ \\
         \hline
          {\tt \href{https://youtu.be/OvzXm38RhFM}{K3}} & $1$ & $8$ & $14$ & $0.9$ &  $100$ & $0.58$   & $0.46$ &  $0.67$ \\
         \hline
         {\tt \href{https://youtu.be/1Fl5PXsHP2M}{K4}} & $1$ & $8$ & $14$ & $0.9$ &  $10$ & $1.83$ & $0.46$  &  $<0.40$ \\
         \hline
         {\tt \href{https://youtu.be/x7d3CIwDqYY}{K5}} & $1$ & $8$ & $14$ & $0.15$ &  $10$ & $1.06$  & $0.46$  & $<0.44$ \\    
         \hline
    \end{tabular}
    \caption{Details of the 3D simulations used in this paper and the main result. Click the name to watch the video on YouTube.  For each model initial parameters are given in the table: name of the model, loops number $N_{\rm loops}$, the inner radius of the disk $r_{\rm in}$, the position of the maximal pressure in the disk $r_{\rm max}$, Kerr parameter $a$, initial magnetic flux in the loop $\Psi_0$ in code units, and initial $H/R$. The last column shows the ratio of the disk mass at the end of the analysis ($10$ hours) to the initial disk mass.} 

    \label{tab:models} 
\end{table*}

 In the next Section 
we describe the numerical models we use. In Section \ref{sec:results}  we discuss how the disk parameters influence the accretion dynamics of the disk and its relation to the observational properties of the QPEs, we also use our simulations to explore the predicted jet power, in order to explain the absence of variable jets in these systems. Finally in Section \ref{sec:discussion} we summarize the results.

\section{Models} \label{sec:model}

We conduct multiple GRMHD simulations utilizing the {\tt HARMPI} code \citep{harmpi}. The simulations are performed in spherical Boyer-Lindquist coordinates $r, \theta, \phi$. The initial conditions for all of the numerical experiments described below involve a standard Fishbone and Moncrief torus \citep{Fishbone1976} surrounding a Kerr black hole. In the code we use units $G=c=1$ and normalize all of our quantities by the black hole mass $M_{\rm BH, 0}$ and the maximal disk density $\rho_0$.  The radial grid is distributed as follows: for $r<400 r_{\rm g}$, we use a logarithmically spaced grid with a constant ${\rm d}r/r$, and for $r>400 r_{\rm g}$, the grid scales as ${\rm d}r/r=4 (\log r)^{3/4}$. The basic resolution is set to $N_r\times N_{\theta} \times N_{\phi} = 384\times 192\times 48$.  We consider two different torii geometries: thin disks with initial $h/r=0.13$ (models {\tt N1, N2}) and thick disks with initial $h/r=0.46$ (models {\tt K1, K2, K3, K4, K5}): see the description of all models in Table~\ref{tab:models}. Since the disks are not too large, we set the outer radial boundary at  $r_{\rm out}=1000 \, r_{\rm g}$. We run one additional simulation of a thick disk with a larger outer boundary $r_{\rm out}=10^5r_{\rm g}$, to check for consistency and verified that the results were unchanged. 

\begin{figure}
    \centering
    \includegraphics[width=0.25\textwidth ,trim= 0cm 0cm 9.45cm 0cm, clip=true]{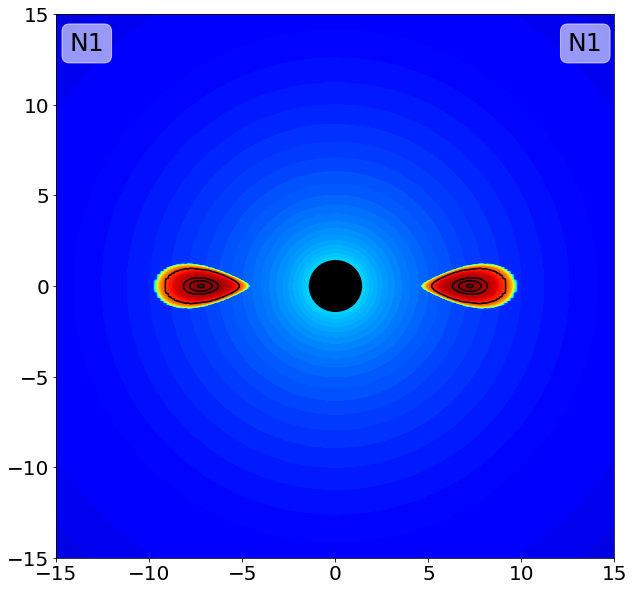}
    \hspace*{-1.35cm} 
    \includegraphics[width=0.25\textwidth ,trim= 9.45cm 0cm 0cm 0cm, clip=true]{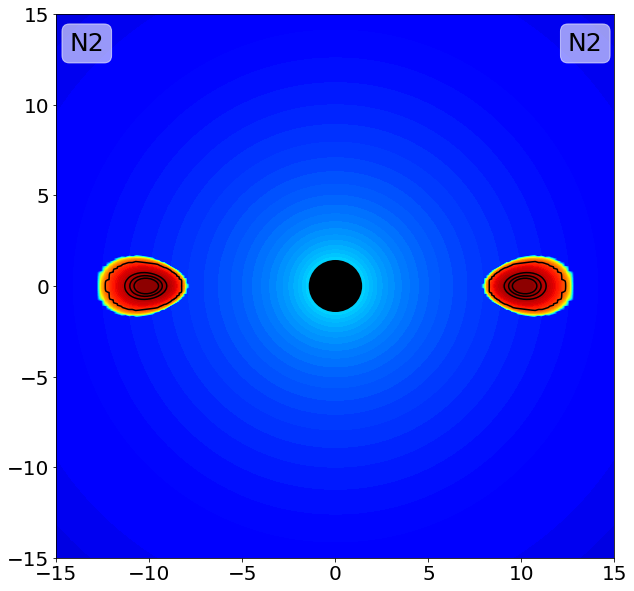}
    
    \caption{The initial density distribution and magnetic field configuration in the thin disk models ({\tt N1},{\tt N2}). Magnetic field lines are marked by solid black lines.} 
    \label{fig:fig1}
\end{figure}

The initial magnetic field in the torus in our simulations contains only a poloidal component and is described by a vector potential $A_{\phi}\rm{\bf{\hat{e}}}_\phi$. We consider two different initial magnetic field configurations: i) a single loop defined by the vector potential
    \begin{equation}
A_{\phi}  = 
\left\{
\begin{array}{lc}
A_0(\rho-\rho_{\rm{cut}})^2 r^4 & \mbox{if $\rho \geq \rho_{\rm{cut}}$}\\
0 &  \mbox{if $\rho <\rho_{\rm{cut}}$}
\end{array} 
\right.      
\end{equation}    
here $\rho_{\rm cut}=0.05\rho_0$ is the threshold density below which the magnetic field vanishes. ii) Multiple dipolar loops set by 
\begin{equation}
A_{\phi}  = 
\left\{
\begin{array}{lc}
A_0(\rho-\rho_{\rm{cut}})^2 r^4\sin(\pi r_{\rm norm}N_{\rm loops}) & \mbox{if $\rho \geq \rho_{\rm{cut}}$}\\
0 &  \mbox{if $\rho <\rho_{\rm{cut}}$}
\end{array} 
\right.      
\end{equation} 
    where $N_{\rm loops}$ is the number of the loops, and 
    \begin{equation}
      \displaystyle  r_{\rm norm}=\displaystyle\frac{\displaystyle r/r^{\rm torus}_{\rm in}-1}{r^{\rm torus}_{\rm out}/r^{\rm torus}_{\rm in}-1},
    \end{equation}
with $r^{\rm torus}_{\rm in}$, $r^{\rm torus}_{\rm out}$ denoting respectively the inner and outer radii of the torus (or more precisely its magnetized section, defined by $\rho > \rho_{\rm cut}$). In our simulations we consider  $N_{\rm loops}=6$. The normalization parameter $A_0$ is chosen to ensure that the required initial ratio of gas to magnetic pressure $\beta=p_{\rm g}/p_{\rm m}\geqslant  \beta_{\rm min}$, where $\beta_{\rm min}$ is set by hand for every simulation, see Table~\ref{tab:models}. This form of magnetic potential gives a more or less constant $\beta$ distribution in the disks, which is important for our ability to resolve the magneto-rotational instability (MRI) \citep{MRI}.

\begin{figure}
    \centering
    
    \includegraphics[width=0.25\textwidth ,trim= 0cm 0cm 9.45cm 0cm, clip=true]{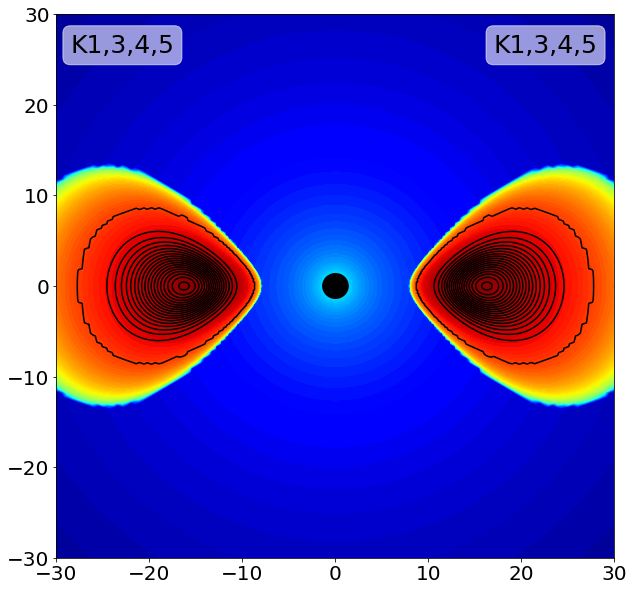}
    \hspace*{-1.35cm} 
    \includegraphics[width=0.25\textwidth ,trim= 9.45cm 0cm 0cm 0cm, clip=true]{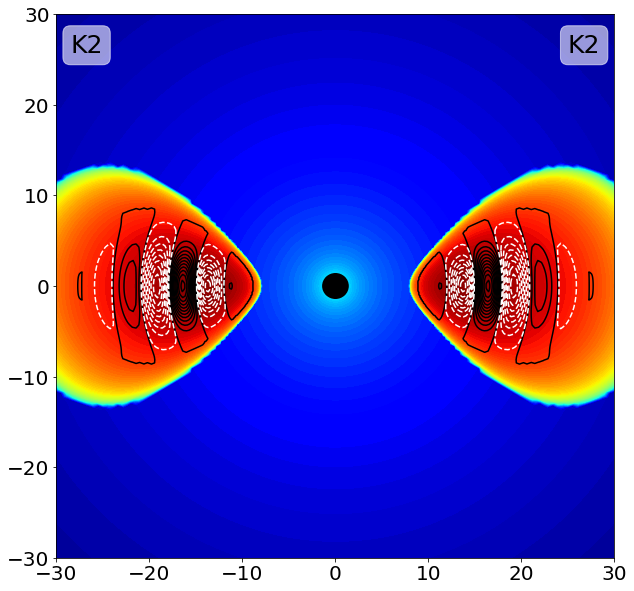}
    
        \caption{
        The initial density distribution and magnetic field configuration in the thick disk models ({\tt K1}-{\tt K5}). Magnetic field lines are marked by solid back lines for clockwise and dashed white lines for counter-clockwise oriented field.} 
    \label{fig:fig2}
\end{figure}

To compare our results with observations we convert the code time units to physical ones, using a characteristic BH mass $M_{\rm BH, 0}=10^6M_{\odot}$ with an associated dynamical time  $t_{\rm d}=4.93M_{\rm BH}/M_{\rm BH,0}$~s.  We do not convert the mass accretion rate from code units to the physical ones because it depends on disk density $\rho_0$, which is a free parameter in our models. We emphasize that the accretion time depends on the disk geometry but not on $\rho_0$. Therefore, $\dot{M}$ can be rescaled by changing $\rho_0$ without affecting the flare duration, or the recurrence time. 
In reality, $\rho_0$ can be inferred from the mean accretion rate (averaged over the orbital period of the donor star), which is determined by the 
rate of mass transfer from the star to the disk and should be $<\dot{M}> \sim 10^{-4}\, {\rm M_{\odot}\, yr^{-1}}$. Note that this quantity merely reflects 
the bolometric luminosity during  quiescent states (i.e., between flaring states), setting a lower limit on the actual accretion rate during a flare.

\section{Results} \label{sec:results}

In this work we focus on three  observed properties of the QPE in GSN~069 and test what type of disks can reproduce them. 
1) The outburst duration $T\sim 1$~hour.
2) A recurrence time of $T_{\rm rec}\sim 10$~hours, which is related to the disk accretion time. 3) The absence of a variable jet, which is connected with the EM energy output from the BH. To test each property we divide our models into three groups:

\begin{itemize}
    \item Models  {\tt N1, N2, K3} are used to understand how the disk morphology, the disk thickness $h/r$ and the position of inner radius $r_{\rm in}$, affects the accretion time and the jet production; 
    \item Models  {\tt K2, K3} are used to understand the effect of the magnetic field configuration;
    \item Models {\tt K1, K3, K4, K5} are used to test the importance of plasma $\beta$ and of the Kerr parameter $a$.
\end{itemize}

For each model we plot the evolution of the mass accretion rate through the horizon normalized by initial disk mass $M_0$:
\begin{equation}\label{eq:Mdot}
\displaystyle\frac{\dot M}{M_0}=-\frac{1}{M_0}\iint\limits_{\theta,\phi}\rho u^r {\rm d}A_{\Omega},
\end{equation}
where $\rho$ is the proper mass density, $u^r$ is the $r$-component of the contravariant 4-velocity, ${\rm d}A_{\Omega}=\sqrt{-g}{\rm d}\theta {\rm d}\phi$ is a solid angle differential area 
and $g$ is the metric determinant. The integral is taken on the BH horizon defined by a sphere with a radius $r\uu{\rm H}=r_{\rm g}(1+\sqrt{1-a^2})$. 

In order to check for the presence of a jet we analyse the jet efficiency  utilizing two different definitions of jet efficiency: accretion jet efficiency $\eta_{a}$ and Blandford-Znajek (BZ) jet efficiency $\eta\uu{\rm BZ}$.  
The accretion jet efficiency  measures the electromagnetic power output from the BH horizon relative to the rest-mass energy flux that crosses the horizon, and is defined as:
\begin{equation}
        \eta_{\rm a}=\displaystyle\frac{F\uu{\rm EM}}{F\uu{\rm M}}  \times 100\%,
\end{equation}
    where $F\uu{M}=\dot Mc^2$ is the rest-mass energy flow through the horizon, and  $F\uu{EM}=\iint [T\uu{EM}]^r_t {\rm d}A\uu{\Omega}$
   is the total electromagnetic (EM) power coming out of the BH horizon with  $[T\uu{EM}]^r_t=b^2u^ru_t-b^rb_t$ denoting the electromagnetic part of the $rt$ component of the stress-energy tensor. 
 The BZ efficiency measures the electromagnetic  power output from the BH horizon relative to the Blandford-Znajek power \citep{BZ77}, which is associated with the initial magnetic flux in a loop.
   \begin{equation}
       \eta\uu{\rm BZ}=\displaystyle\frac{F\uu{EM}}{F\uu{BZ_0}}\times 100\%,
   \end{equation}
   where $F\uu{BZ_0}=\displaystyle \frac{1}{24\pi^2 c}\Omega^2\uu{BH}\Psi_0^2$ , $\Psi_0$ is the initial magnetic flux in the loop \citep{Tchekhovskoy2012}, and $\Omega\uu{BH}=\displaystyle\frac{ac}{2r\uu{H}}$ is the BH angular frequency. In the case of multiple loops $\Psi_0$ is related to the loop with the maximal magnetic flux. The first efficiency shows how luminous the jet is relative to the accretion power, while the second efficiency measures how effective the magnetic flux accumulation on the BH horizon is.

\subsection{Disk geometry and position}

\begin{figure}
    \centering
    \includegraphics[width=0.5\textwidth]{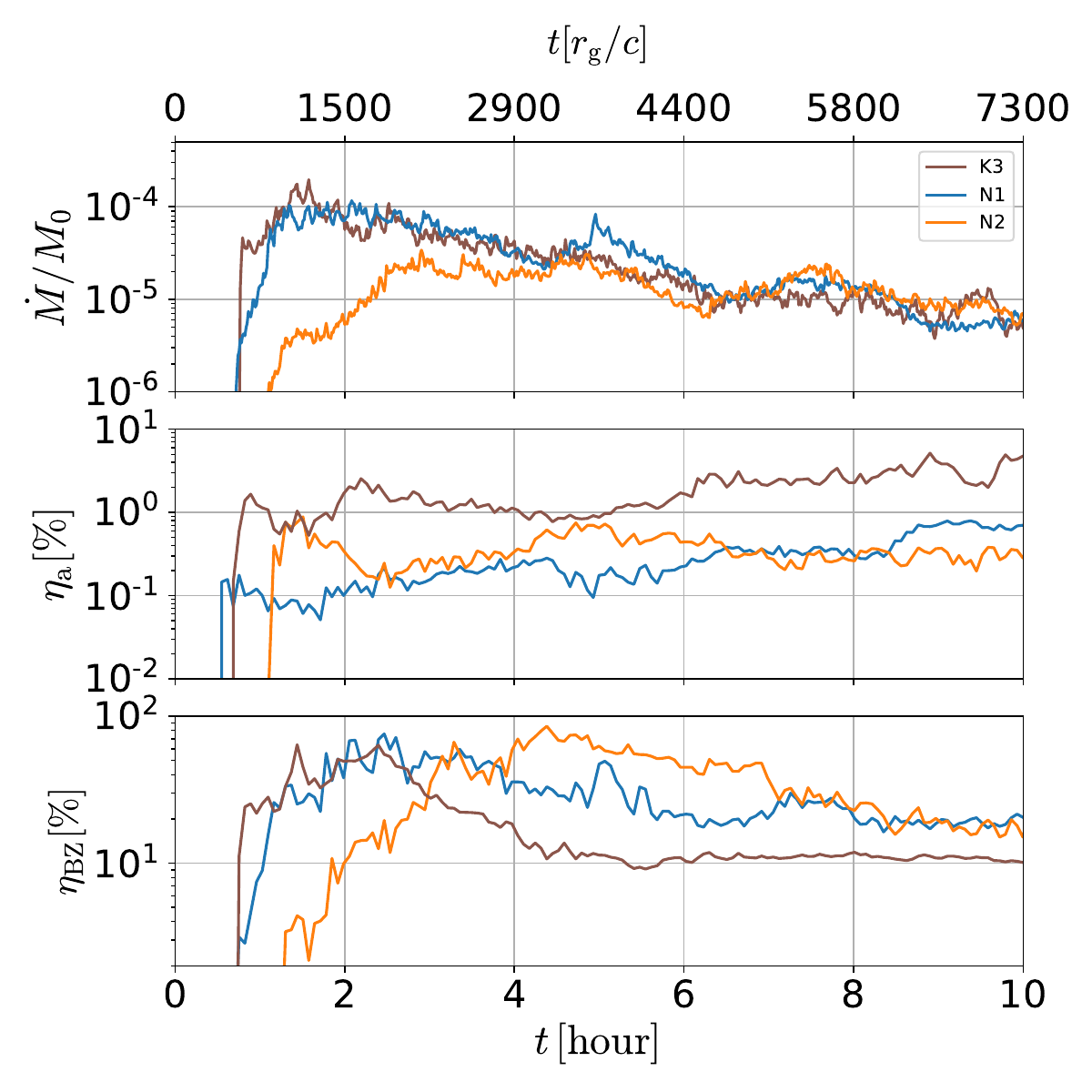}
    \caption{The evolution of mass accretion rate normalized by the initial disk mass and of the jet efficiencies, all measured at the BH horizon, in models {\tt N1, N2, K1}. The time is presented in hours, appropriate for a BH mass of $10^6M_\odot$ (lower $x$ axis), and in natural units of $r_{\rm g}/c$ (top $x$ axis).
    }
    \label{fig:fig3}
\end{figure}

\begin{figure}
    \centering
    \includegraphics[width=0.5\textwidth]{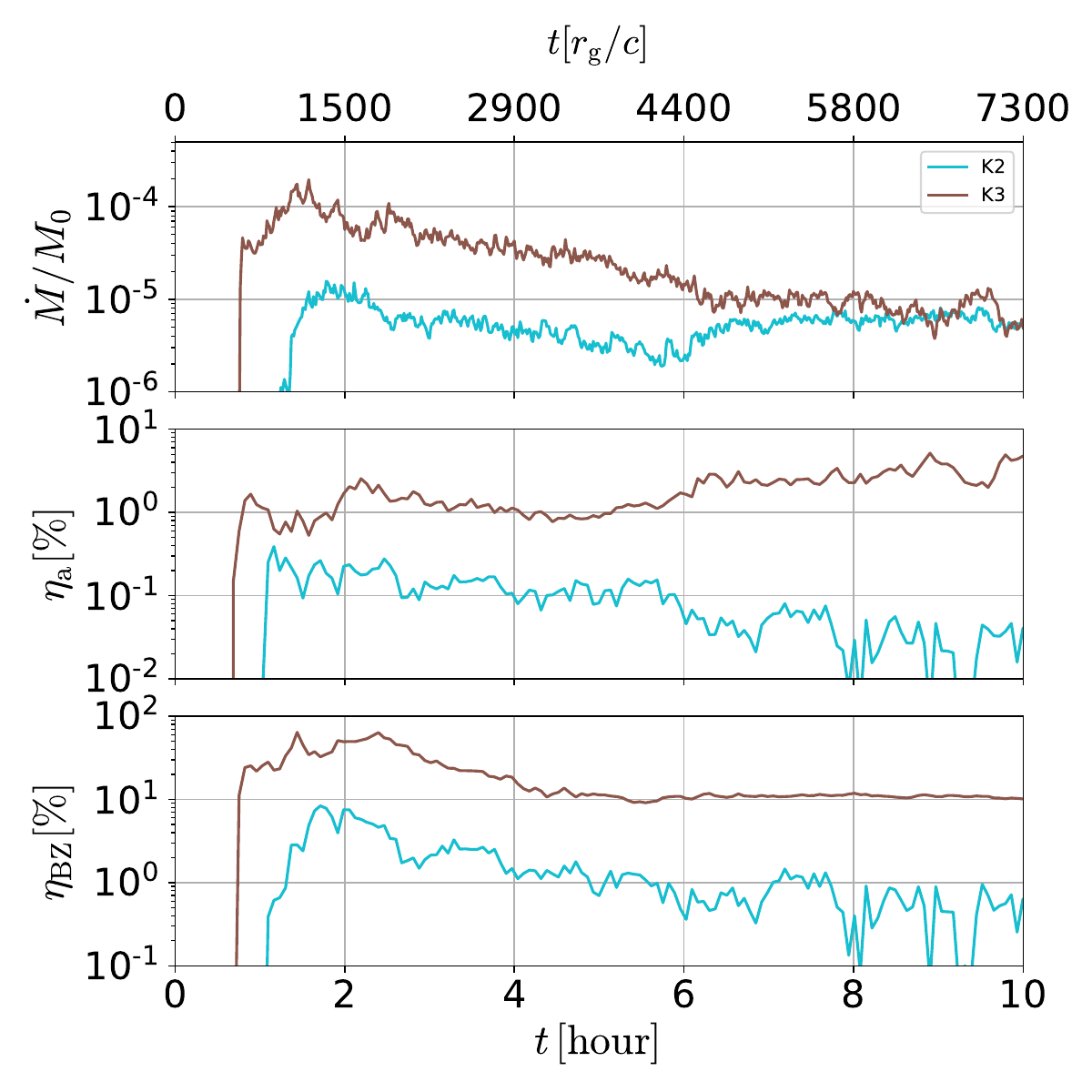}

    \caption{Same as Fig.~\ref{fig:fig3} for the thick disk models {\tt K2, K3}.}
    \label{fig:fig4}
\end{figure}

In the first set of models we examined how the
disk geometry and its inner radius affect the observational parameters.  For this we considered three models: {\tt N1, N2, K3}. The first two models have thin disks with initial $h/r\sim 0.13$, while the {\tt K3} model features a thick disk with an initial $h/r\sim 0.46$.  In addition, the inner disk radius in model {\tt N1} is half the inner radius in models {\tt N2} and {\tt K3}.
All three disks have the same initial plasma $\beta = 100$, so we expect the jet efficiency to be affected by the disk geometry alone.

The results of the simulations are shown in Fig.~\ref{fig:fig3}. Two of the three models show similar  evolution of mass accretion rate: a thin disk with a smaller inner radius ({\tt N1}) and a thick disk with a larger inner radius ({\tt K3}). The mass accretion rate peaks about 1.5 hours after the beginning of the simulation and then drops by an order of magnitude over 6 hours. The similarity means that it is hard to say just from the inferred mass accretion rate whether we see a thin, dense disk close to a black hole or a thick, more tenuous disk that is formed further away from the black hole.
Comparison between the behaviour of $\dot M$ in models {\tt N1} and {\tt N2} having the same disk thickness shows that the position of the inner disk radius affects mostly the onset time of the accretion, where accretion starts later in disks with larger inner radii.

 The BZ jet efficiency $\eta\uu{\rm BZ}$  is quite the same in the thin disks {\tt N1} and {\tt N2},  having a value of about $100\%$, implying that magnetic flux stored initially in the disks is efficiently converted into jet's Poynting flux. In the thick disk simulation {\tt K3} the BZ jet efficiency also reaches about $100\%$ but then quickly drops to $\sim10\%$. However, the accretion jet efficiency $\eta_{\rm a}$ in the thin disk models is only sub-percent, while in the thick disk {\tt K3} it is an order of magnitude higher. The reason for that is that the disks in all models do not have enough magnetic flux\footnote{It is possible to increase the total flux by increasing the disk volume or decreasing plasma-$\beta$, but here we study the influence of disk geometry  without changing the other parameters.} initially stored in the disk and they cannot reach a MAD state, which is signified by an efficient energy extraction from the BH. 

We therefore conclude that the position of the inner radius of the disk affects the beginning of the accretion, but thick disk can mimic a thin dense disk that is located closer to a black hole.

\subsection{Magnetic field configuration}

To study the effect of magnetic field on the QPE  we compare between two thick disk models, {\tt K2} and {\tt K3} (Fig.~\ref{fig:fig4}). 
The disk in the model {\tt K2} has six magnetic loops of alternate polarity, representing a more random magnetic field configuration, while {\tt K3} has a single loop.
The initial flux in each loop of model {\tt K2} is about the same and is $2.5$ times smaller than the magnetic flux in the model {\tt K3}. We do not consider thin disks here since resolving the MRI in such disks, which is necessary to obtain a physical evolution of the disk magnetic field, requires resolutions much higher than our computational capabilities.

Accretion onto the BH in the case of a single loop ({\tt K3}) starts sooner and reaches a rate $\sim 10$ times higher than in the case of random field ({\tt K2}). 
The main reason for this difference is that in model {\tt K2} the radial component of the magnetic field is much smaller than in {\tt K3}. This reduces the effective viscosity in the disk (the viscosity depends on Maxwellian stresses $T_{r\phi} \sim B_rB_{\phi}$). 
A second reason originates from reconnection in the small scale field of model {\tt K2}, that starts soon after the onset of the simulation and prevents an efficient additional buildup of magnetic field by MRI. As a result the disk of model {\tt K2}, features a significantly lower accretion rate than the disk of model {\tt K3} despite the fact that both disks had a similar initial magnetic flux.   The mass accretion rate in model {\tt K2} remains roughly constant throughout the entire simulation.

Another outcome of the complex magnetic field topology is a smaller magnetic flux buildup on the BH horizon, with a peak flux $\sim 1/6$ of the peak flux in model {\tt K3} , resulting in a lower jet efficiency by an order of magnitude (Fig.~\ref{fig:fig4}). So we can conclude that the small-scale magnetic field simulations produce a smaller mass accretion rate, a lower magnetic flux through the horizon and a low jet efficiency.

\subsection{Plasma $\beta$ and Kerr parameter $a$}

\begin{figure}
    \centering
    \includegraphics[width=0.5\textwidth]{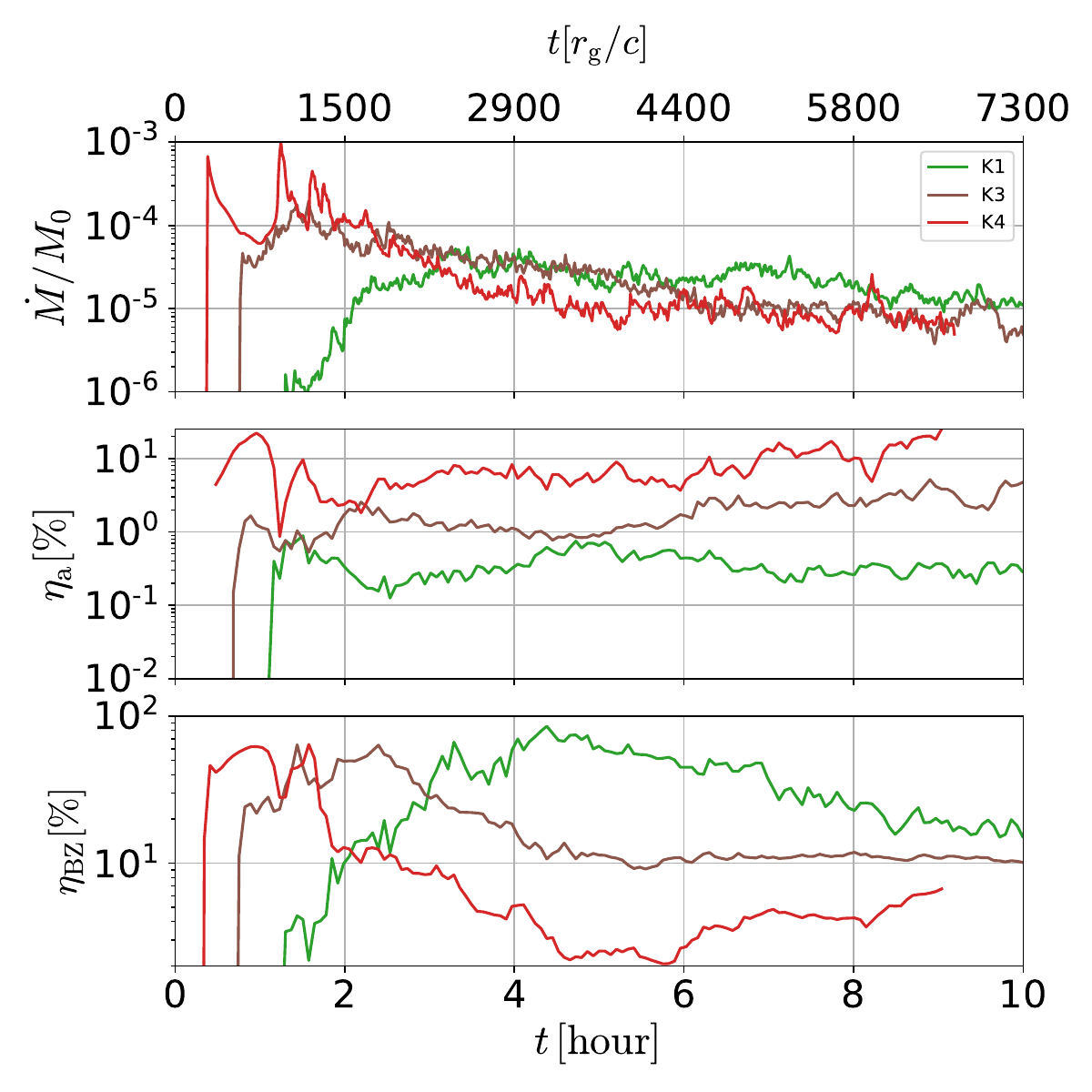}
    \caption{Same as Fig.~\ref{fig:fig3} for  models {\tt K1, K3, K4}.}
    \label{fig:fig5}
\end{figure}

\begin{figure}
    \centering
    \includegraphics[width=0.5\textwidth]{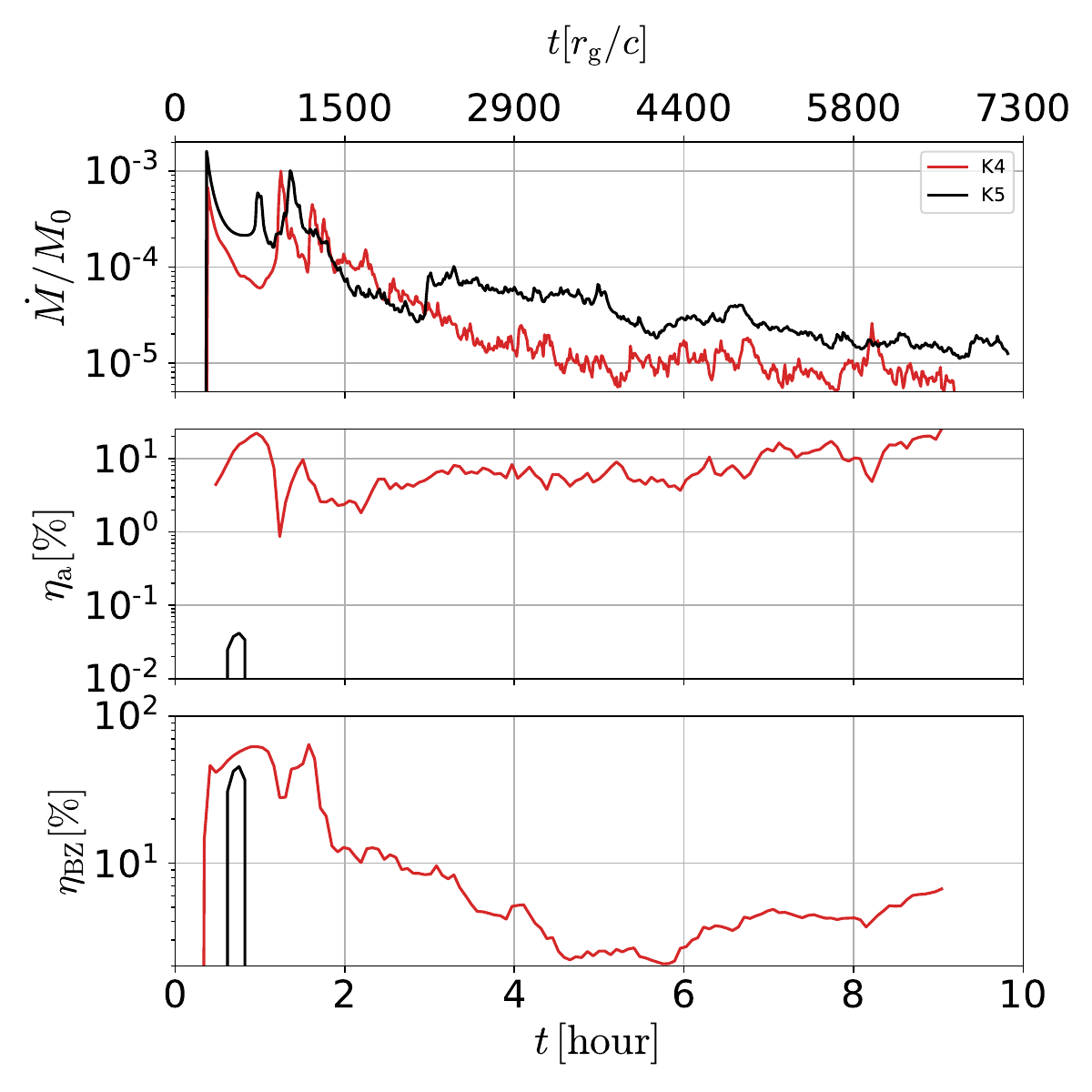}
    \caption{Same as Fig.~\ref{fig:fig3} for models {\tt K4, K5}.}
    \label{fig:fig6}
\end{figure}

To study the effect of initial magnetization on the observational parameters we compare the evolution of three thick disk models {\tt K1, K3, K4}  with initial  $\beta=1000$, $100$, and $10$, respectively. All these models have the same geometry and magnetic field configuration. 
The evolution of the compared parameters is shown in
Fig.~\ref{fig:fig5}. We see that in more magnetized disks, accretion starts sooner and the magnetic flux on the horizon accumulates faster and more efficiently. This determines the evolution of the BZ efficiency, which reaches a peak of $\sim70\%$ faster in the highly magnetized disk models, {\tt K3} and {\tt K4}, than in model {\tt K1}. 
 The mass accretion rate increases rapidly in the disks with $\beta=10-100$, but it decreases by a factor of $10$ within $2-6$ hours,  becoming comparable to the less magnetized model {\tt K1}. The accretion jet efficiency is higher for the highly magnetized disks and reaches a value of $\sim 10\%$ in model {\tt K4} (with $\beta=10$), while in model {\tt K1}, with $\beta=1000$, it remains below $1\%$ as it has a lower initial magnetic flux in the disk , which is also converted into the jet's Poynting flux slower than for the other thick disk models. 
 
Low accretion jet efficiency is seen, so far, only in the multiple loops case (model {\tt K2}). However, low accretion jet efficiency can be reproduced in systems with a slowly rotating black hole as well. 
As shown by \citet{Lowell2023}, the accretion jet efficiency $\eta\uu{a}$ (in their paper it is called as the electromagnetic outflow efficiency) drops when $a<0.2$. 
To test that compared two thick disk models {\tt K4} and {\tt K5}, which differ only in their Kerr parameter: {\tt K4} has a fast rotating BH ($a=0.9$) and {\tt K5} has a slower-rotating BH ($a=0.15$). The results are shown in Fig.~\ref{fig:fig6}. The mass accretion rate is the same for both models, showing high rates in the first two hours and dropping by an order of magnitude at longer times.

However, the accretion and BZ jet efficiencies in the slowly rotating BH are practically zero, making model {\tt K5} an ideal candidate to explain the lack of a variable jet in the QPEs.

\subsection{Recurrence time}

\begin{figure}
    \centering
    \includegraphics[width=0.5\textwidth]{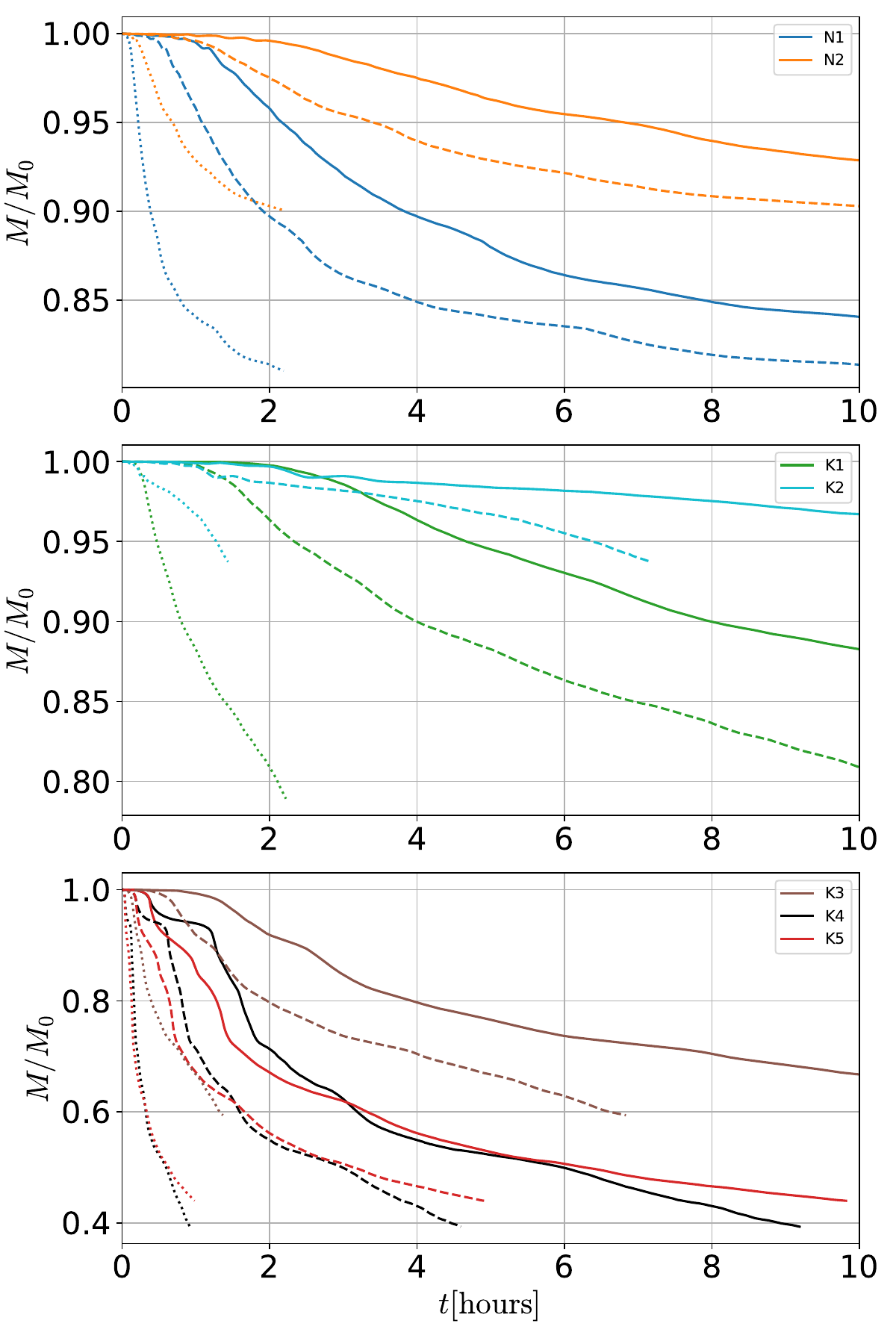}
 
    \caption{The ratio of total disk mass to the initial disk mass versus time in the different simulations. For each model we show the evolution of the disk mass fraction for three different BH masses: solid line ($M_{\rm BH}=10^6 M_{\odot}$), dashed line ($M_{\rm BH}=5\cdot 10^5 M_{\odot}$) and dotted line ($M_{\rm BH}=10^5 M_{\odot}$). }
    \label{fig:fig7}
\end{figure}

Last, in order to explain the observed recurrence time of 10 hours, we require that most of the disk mass will be accreted within this time. If a significant fraction of the initial disk mass is left after $10$ hours, 
additional mass that will fall on the disk in the following cycles will result in a continuous growth of the disk mass within a few orbital periods, increasing the quiescent emission and erasing the QPE signature.

In Fig.~\ref{fig:fig7} we plot the current to initial disk mass ratio versus time for three different BH masses: $10^6M_{\odot}$, $5\cdot 10^5M_{\odot}$ and $10^5M_{\odot}$ with solid, dotted and dashed lines respectively.  
We measure the disk mass as the mass of the bounded material in the simulation. The matter is gravitationally bound if
\begin{equation}
h\cdot u_t<-1,
\end{equation}
where $h=1+\gamma u_g/\rho$ is the specific enthalpy, $\gamma=4/3$ is the adiabatic index, $u_g$ is the thermal energy density and $u_t$ is the temporal covariant component of 4-velocity (equal to minus energy at infinity). 
The comparison is done for all models in this work.
Only models {\tt K3, K4} and {\tt K5} featuring thick, relatively highly magnetized disks, show a significant drop in the disk mass  within the 10 hours cycle time and BH mass range appropriate for GSN 069. The mass loss in model {\tt K3} is marginal compared with the other two models, owing to a higher initial plasma $\beta$. Note that in all thick disk models a significant contribution to the mass loss comes from strong disk winds. This is at odds with the thin disk models ({\tt N1, N2}), in which the mass loss from winds is insignificant. Thus, despite the fact that models 
{\tt N1} and {\tt K3} show similar accretion history (see Fig. \ref{fig:fig3}), model {\tt N1} has a higher disk mass at the end of the cycle, 
making it incompatible with the observational constraint of low quiescent emission.


\section{Discussion} \label{sec:discussion}

In this work we explored the possibility that the QPEs observed in GSN~069 occur due to regular cycles of mass buildup in a small accretion disk around the SMBH. We focused on three main observables: the flare duration, the time between outbursts and the lack of evidence for a variable jet emission. All three thick,
highly magnetized disk models {\tt K3-K5} can account for the required observables within some margins. Out of the three, model {\tt K5}, featuring an outburst behaviour at the beginning of the cycle, a significant mass loss by the end of the cycle and no apparent jet activity seems the most reasonable candidate. Model {\tt K4} shows similar accretion properties to model {\tt K5}, however with a higher jet accretion efficiency of $\eta\uu{a}\sim10\%$. 
In order to avoid emitting a significant, variable radio emission the jets in model {\tt K4} have to accelerate freely to a distance $r\gg c\cdot10~\rm{hours}\sim10^4M_6^{-1} {\rm r_g}$ before they can be decelerated due to interaction with the ambient medium and radiate their energy.
Model {\tt K3} satisfies two out of the three observational constraints. The disk accretes a significant part of its mass within $10$ hours, especially if the black hole mass is lower than $10^6M_{\odot}$, and it has a small jet efficiency. However, the lack of a clear outburst behaviour at the beginning of the cycle makes it less plausible candidate. 


The recurrence time of $10$~hours requires a combination of a thick disk with $\beta \lesssim 100$ and a large scale magnetic field (as can be seen in Models {\tt K3-K5}). This combination allows the MRI to operate faster and produce a higher effective disk viscosity. These conditions could be difficult to fulfill if the disk is formed from partial disruption of regular stars. There is no strong evidence that the magnetic field in the stellar matter has such high magnetization and regular structure. It is more likely that in this case the initial magnetic field in the disk has randomly distributed small-scale magnetic field, which leads to relatively slow accretion, as can be seen in model {\tt K2}. The problem may be resolved if we assume that the matter forming the disk originates from a stellar wind. In stellar winds the gas pressure drops faster than the magnetic pressure, thus the plasma $\beta$ should be small.  
Such a scenario can be realized if we consider a binary system of two S-stars with a binary period of $\sim 10$~hours orbiting a BH. These stars typically have strong stellar winds. Interacting winds from the binary system can provide the BH with a low angular momentum, strongly magnetized flow of matter, variable on the timescale of the binary period. 
 The main role of stellar winds from S-stars in forming the accretion  disk in the Galactic center is supported by numerical simulations  by  \citet{Calderon2021}.

Last, we note that in our simulations we consider ADAF-like disks, without cooling. Cooling reduces the disk thickness and decreases the plasma $\beta$ by lowering the gas pressure within the disk. This should have two opposite effects. On one hand, the mass accretion rate in thin disks is expected to be smaller\footnote{According to the standard disk model \citep{SS73}, thin disks accrete slower because the mass accretion rate $\dot M$, is proportional to the radial velocity $v_r$, which depends on the thickness as $v_r= \alpha v_{\rm K}(h/r)^2$.}. On the other hand the small $\beta$ increases the viscosity parameter $\alpha$ in the disk and thus the mass accretion rate.  Recent work by \citet{Scepi2023} shown that the accretion rate in thin disks is higher than what predicted by analytical models. Therefore, the range of accretion disk parameters explaining QPEs may be broader than the one predicted here. 
The effect of cooling will be investigated in a separate work.

\begin{acknowledgements}
      We thank Pavel Abolmasov and Matthew Green for helpful discussions. 
      This work was supported by a grant from the Simons Foundation (00001470) to AL.
\end{acknowledgements}

\bibliographystyle{aa}
\bibliography{mybib}

\end{document}